\documentclass[english,prd,eqsecnum,nofootinbib,floatfix]{revtex4}
\usepackage[T1]{fontenc}
\usepackage[latin9]{inputenc}
\usepackage{amsmath}
\usepackage{amssymb}
\usepackage{esint}

\makeatletter
\@ifundefined{textcolor}{}
{%
 \definecolor{BLACK}{gray}{0}
 \definecolor{WHITE}{gray}{1}
 \definecolor{RED}{rgb}{1,0,0}
 \definecolor{GREEN}{rgb}{0,1,0}
 \definecolor{BLUE}{rgb}{0,0,1}
 \definecolor{CYAN}{cmyk}{1,0,0,0}
 \definecolor{MAGENTA}{cmyk}{0,1,0,0}
 \definecolor{YELLOW}{cmyk}{0,0,1,0}
 }

\usepackage{amsfonts}
\providecommand{\U}[1]{\protect\rule{.1in}{.1in}}
\newcommand{\BOX}{\hbox {$\sqcap$ \kern -1em $\sqcup$}}\newcommand{\be}{\begin{equation}}\newcommand{\ee}{\end{equation}}\newcommand{\ba}{\begin{eqnarray}}\newcommand{\ea}{\end{eqnarray}}\newcommand{\ban}{\begin{eqnarray*}}\newcommand{\bea}{\begin{eqnarray}}\newcommand{\eea}{\end{eqnarray}}\newcommand{\ean}{\end{eqnarray*}}\newcommand{\barr}{\begin{array}}\newcommand{\earr}{\end{array}}

\makeatother

\usepackage{babel}

\makeatother

\usepackage{babel}

\makeatother

\usepackage{babel}

\makeatother

\usepackage{babel}

\makeatother

\usepackage{babel}

\makeatother

\usepackage{babel}

\begin{document}
KCL-PH-TH/2010-28

\title{Finsler metrics and CPT}

\author{Sarben Sarkar }

\affiliation{King's College London, Department of Physics, Strand, London WC2R
2LS, UK. }
\begin{abstract}
The role of Finsler-like metrics in situations where Lorentz symmetry
breaking and also CPT violation are discussed. Various physical instances
of such metrics both in quantum gravity and analogue systems are discussed.
Both differences and similarities between the cases will be emphasised.
In particular the medium of D-particles that arise in string theory
will be examined. In this case the breaking of Lorentz invariance,
at the level of quantum fluctuations, together with concomitant CPT
in certain situations will be analysed. In particular it will be shown
correlations for neutral meson pairs will be modified and a new contribution
to baryogenesis will appear.
\end{abstract}
\maketitle

\section{Introduction \label{sec1}}

Paul Finsler in 1918 wrote a thesis, {}``Über Kurven und Flächen
in Allgemeiner Rämen'', which is a generalisation of Riemannian manifolds.
It remained a topic in the province of pure mathematics until quite
recently. However, within physics, it has been realised more recently
that a framework more general than pseudo-Riemannian geometry is required
if causal relations, which are more complicated that special relativity,
are involved. It is interesting that this realisation has sprung up
in somewhat diverse contexts. We shall be primarily interested in
the particular context known as space-time D foam, where it appears
in a more generalised way. In general, the mathematical structure
that is unveiled in physical applications, is different from the original
Finsler theory in that the associated norm is not positive definite;
hence, following physics tradition, it can be dubbed pseudo-Finsler.
However, in the foamy context, a stochastic aspect to the norm is
introduced.

At each point $x$ in a (standard) Finsler manifold $M$ \cite{finsler}
there is a norm on the tangent space $TM_{x}$ which has not been
induced by an inner product. Rather the norm itself induces an inner
product. However this inner product is not parametrised by points
in $M$ but by directions in $TM$. Specifically a Finsler norm $F$
on $TM$ is a smooth mapping on $TM\setminus\left\{ 0\right\} \left(\equiv\cup\left\{ T_{x}M\setminus\left\{ 0\right\} :\, x\in M\right\} \right)$.
Now $F\mid_{T_{x}M}:TM\rightarrow\left[0,\infty\right)$is such that
$F\mid_{T_{x}M}\left(\equiv f_{x}\right)$ is homogeneous of degree
$1$ and, for all $y$ in $TM\setminus\left\{ 0\right\} $, a form
$g_{y}:T_{x}M\times T_{x}M\rightarrow\mathbb{R}$ can be defined by
\[
g_{y}:\left(u,v\right)\rightarrow\frac{1}{2}\frac{\partial^{2}\left\{ \left[f_{x}\left(y+pu+qv\right)\right]^{2}\right\} }{\partial p\partial q}\left|_{p=q=0}.\right.\]
 The form $g_{y}$ is bilinear and required to be positive defnite.
A Riemannian manifold $\left(M,G\right),$$G$ being the metric, trivially
can be recast as a Finsler manifold with\[
F\left(x,y\right)=\sqrt{G_{x}\left(y,y\right)}.\]
 Here $x$ is a point in $M.$

\subsection{The stochastic Finsler metric}

String theory currently attracts much attention in that quantum states
of gravitons are part of its spectrum together with other states which
are required for the standard model of particle physics. The discovery
of new solitonic structures in superstring theory~\cite{polch2}
has dramatically changed the understanding of target space structure.
These new non-perturbative objects are known as D-branes and their
inclusion leads to a scattering picture of space-time fluctuations.
Typically open strings interact with D-particles and satisfy Dirichlet
boundary conditions when attached to them. Closed strings may be cut
by D-particles. D-particles are allowed in certain string theories
such as bosonic, type IIA and type I and here we will here consider
them to be present in string theories of phenomenological interest.
Furthermore even when elementary D particles cannot exist consistently
there can be effective D-particles formed by the compactification
of higher dimensional D branes. Moreover D particles are non-perturbative
constructions since their masses are inversely proportional to the
the string coupling $g_{s}$. The study of D-brane dynamics has been
made possible by Polchinski's realization that such solitonic string
backgrounds can be described in a conformally invariant way in terms
of world sheets with boundaries. On these boundaries Dirichlet boundary
conditions for the collective target-space coordinates of the soliton
are imposed. Heuristically, when low energy matter given by a closed
(or open) string propagating in a $\left(D+1\right)$-dimensional
space-time collides with a very massive D-particle embedded in this
space-time, the D-particle recoils as a result. Since there are no
rigid bodies in general relativity the recoil fluctuations of the
brane and their effectively stochastic back-reaction on space-time
cannot be neglected.Based on these considerations, a model for a supersymmetric
space-time foam has been suggested in \cite{emw} . The model is based
on parallel brane worlds (with three spatial large dimensions), moving
in a bulk space time which contains a {}``gas'' of D-particles.
The number of parallel branes used is dictated by the requirements
of target-space supersymmetry in the limit of zero-velocity branes.
One of these branes represents allegedly our Observable Universe.
As the brane moves in the bulk space, D-particles cross the brane
in a random way. From the point of view of an observer in the brane
the crossing D-particles will appear as flashing on and off space-time
defects, that is microscopic space-time fluctuations. This will give
the four-dimensional brane world a {}``D-foamy'' structure.

Interactions in string theory, at the present moment, are not treated
systematically as a second quantised formalism is lacking. An important
consistency requirement of first quantised string theory is conformal
invariance, which determines the space-time dimension and/or structure.
On the brane there are closed and open strings propagating. Each time
these strings {}``meet'' a D-particle, there is a possibility of
being attached to it. The entangled state causes a back reaction onto
the space-time, which can be calculated perturbatively using logarithmic
conformal field theory formalism~\cite{kmw}. Some details are reviewed
in Appendix A.

Let $M_{QG}\left(=\frac{M_{S}}{g_{s}}\right)$ be the quantum gravity
mass scale with $g_{s}$being the string coupling and $M_{S}$ the
string mass scale. Even at low energies $E$, decoherence effects
of this foam can have magnitude $O\left(\left[\frac{E}{M_{QG}}\right]^{n}\right)$,
where $n$ is model-dependent with values. Usually $M_{QG}$ is taken
to be $M_{P}$, the Planck mass. Hence, if $M_{QG}$ can be decreased
by many orders of magnitude, then the decoherence effect is duly enhanced.
This decoherence is due to the topologically non-trivial interactions
of the D-particles with strings. We shall see that in some models,
involving large extra dimensions, $M_{QG}$ can reduce to TeV scale.
Such models were devised to deal with the hierarchy of scales between
the Planck and electro-weak scales. Indeed, in the absence of large
extra dimensions, at the heuristic level within the context of microsocpic
black holes, two quarks inside a proton can be absorbed by a virtual
black hole with a re-emission into an antiquark and lepton. An estimate
for the probability of this process can be given in terms of the quantum
chromodynamics scale $(\Lambda_{QCD})$ and $M_{QG}$. Typically the
proton size is $\Lambda_{QCD}^{-1}$, and so, the probability of two
quarks to come within a distance $l_{P}$ within the lifetime of the
black hole, is $O\left(\left(\frac{\Lambda_{QCD}}{M_{QG}}\right)^{4}\right)$.
The constraint on proton lifetime being $10^{33}\mathrm{yr}$ or longer
imples that $M_{QG}>10^{16}\mathrm{GeV}$\cite{Anchordoqui:2006xv}.
Fast baryon decay can be inhibited by separating the quark and lepton
in a large extra dimension, resulting in overlaps of wavefunctions
being reduced and a smaller $M_{QG}$. A model\cite{ArkaniHamed:1998rs}
which can produce $M_{QG}$ of TeV size has been proposed and involves
a $4+n$ dimensional Universe with structure $M_{4}\times K_{n}$,
where $M_{4}$ is a $3+1$ dimensional domain wall ( brane) and $K_{n}$is
a flat $n$-dimensional manifold with extra dimensions of size $R$.
In the model, all standard model interactions in particle physics
would be confined on $M_{4}$, but not gravity. This leads to an $M_{QG}$
satisfying:\[
M_{QG}=\frac{1}{2\pi R}\left(2\pi RM_{P}\right)^{\frac{2}{n+2}}.\]
 For example if $n=2$ and $R=0.1\mathrm{mm}$ then $M_{QG}=10$TeV.

Using the brane model \cite{Dfoam} for space-time fluctuations one
can obtain the following expression for the induced space-time distortion
as a result of the D-particle recoil, in the weakly coupled string
limit, which will be appropriately used in : \begin{equation}
g_{ij}=\delta_{ij},\, g_{00}=-1,g_{0i}=\varepsilon\left(\varepsilon y_{i}+u_{i}t\right)\Theta_{\varepsilon}\left(t\right),\; i=1,\ldots,D\label{recmetr}\end{equation}
 where the suffix $0$ denotes temporal (Liouville) components and
\begin{eqnarray}
\Theta_{\varepsilon}\left(t\right) & = & \frac{1}{2\pi i}\int_{-\infty}^{\infty}\frac{dq}{q-i\varepsilon}e^{iqt},\label{heaviside}\\
u_{i} & = & \left(k_{1}-k_{2}\right)_{i}\;,\end{eqnarray}
 with $k_{1}\left(k_{2}\right)$ the momentum of the propagating closed-string
state before (after) the recoil; $y_{i}$ are the spatial collective
coordinates of the D particle and $\varepsilon^{-2}$ is identified
with the target Minkowski time $t$ for $t\gg0$ after the collision~\cite{kmw}
(see Appendix A). These relations have been calculated for non-relativistic
branes where $u_{i}$ is small and require the machinery of logarithmic
conformal field theory. Now for large $t,$ to leading order,

\begin{equation}
g_{0i}\simeq\overline{u}_{i}\equiv\frac{u_{i}}{\varepsilon}\propto\frac{\Delta p_{i}}{M_{P}}\label{recoil}\end{equation}
 where $\Delta p_{i}$ is the momentum transfer during a collision
and $M_{P}$ is the Planck mass (actually, to be more precise $M_{P}=M_{s}/g_{s}$,
where $g_{s}<1$ is the (weak) string coupling, and $M_{s}$ is a
string mass scale); so $g_{0i}$ is constant in space-time but depends
on the energy content of the low energy particle.

\subsection{Decoherence and non-unitary evolution}

The recoilng D-particle background leads to a fluctuating background.
In string theory for fixed backgrounds to be consistent with a classical
space-time interpretation the central charge needs to have the critical
value. Quantum scattering off the D-particle excites an open string
state which attaches itself to the D-particle. Closed-to-open string
amplitudes determine these excitation and emission processes. To allow
in principle for fluctuating backgrounds it is necessary to deform
the conformal points through vertex operators$V_{g^{I}}$ associated
with background fields$g^{I}$.The world-sheet action $S_{\sigma}$
has a structure \[
S_{\sigma}=S^{*}+g^{I}\int_{\Sigma}V_{I}\left(\Xi\right)\, d^{2}\xi,\]
 which is a (weakly) non-conformal deformation of $S^{*}$, the conformal
action,$V_{I}$ is a vertex (logarithmic conformal field theory) operator
associated with the D-particle, $\Xi$ are target space matter fields,
$\Xi$ is the usual holomorphic world sheet co-ordinate, and $\Sigma$is
the world-sheet surface. In our case the relevant $g^{I}=u_{i}$ with
$i=1,2,3$, if the D-particle recoil is confined to the D3 brane on
which matter strings propagate. Since we are interested in the dynamics
of matter, we will consider the reduced density matrix of matter $\rho_{M}$
on a fixed genus world-sheet.

The renormalization group for field theories in two dimensions has
properties which will have an interesting reinterpretation for string
theories \cite{shore,zam}. In theory space, i.e. in the infinite
dimensional space of couplings $\left\{ g^{I}\right\} $, for all
renormalizable two-dimensional unitary quantum field theories, there
exists a function $c(\{g^{I}\})$ which has the following properties:
\begin{itemize}
\item $c(\{g^{I}\})$ is non-negative and non-increasing on renormalization
group flows towards an infrared fixed point
\item the renormalization group fixed points are also critical points of
$c(\{g^{I}\})$
\item critical value of $c(\{g^{I}\})$ is the conformal anomaly
\end{itemize}
More precisely in terms of a renormalization group flow parameter
(scale) $t$ \begin{equation}
\frac{d}{dt}c(\{g^{I}\})=\beta^{I}G_{IJ}\beta^{J}\label{zam}\end{equation}
 where the renormalization group $\beta$ function is defined by \[
\beta^{I}=\frac{d}{dt}g^{I}(t)\]
 and $G_{IJ}$ is referred to as the Zamolodchikov metric. $G_{IJ}$is
negative definite and is the matrix of second derivatives of the free
energy. In the approach to perturbative string theory based on conformal
field theory on the world sheet , it has long been thought that the
time evolution of the string backgrounds and world sheet renormalization
group flows are connected \cite{gutperle}\cite{EllisErice}. In non-critical
strings conformal invariance is restored by gravitational dressing
with a Liouville field $\varphi$ (which can be viewed as a local
world sheet scale) so that (for convenience now dropping the index
$I$) \begin{equation}
\int d^{2}zgV_{g}(\Xi)\rightarrow\int d^{2}zge^{\alpha_{g}\varphi}V_{g}(\Xi).\label{dressing}\end{equation}
 However $V_{g}$ has a scaling dimension $\alpha_{g}$ to $O(g)$
with $\alpha_{g}=-g\textsl{C}_{ggg}+\ldots$ and $\textsl{C}_{ggg}$
is the expansion coefficient in the operator product expansion of
$V_{g}$ with itself. In a small $g$ expansion \[
\int d^{2}zgV_{g}(\Xi)\rightarrow\int d^{2}zgV_{g}(\Xi)-\int d^{2}zg^{2}\textsl{C}_{ggg}\varphi V_{g}(\Xi).\]
 Scale invariance is restored by defining a renormalized coupling
$g_{R}$ \begin{equation}
g_{R}=g-\textsl{C}_{ggg}\varphi g^{2}\label{dressing2}\end{equation}
 The local scale interpretation of $\varphi$ is clearly consistent
with the renormalisation group $\beta$ function that we have defined
earlier. On noting that $\gamma_{\alpha\beta}=e^{\varphi}\widehat{\gamma}_{\alpha\beta}$
where $\widehat{\gamma}_{\alpha\beta}$ is a fiducial metric, the
integration over world sheet metrics $\gamma_{\alpha\beta}$ ( in
the Polyakov string action) implies an integration over $\varphi$
. In this way $\varphi$ becomes a dynamical variable with a kinetic
term. For matter fields with central charge $c_{m}>25$ the signature
of this term is opposite to the kinetic terms for the fields $\Xi$
and it has been suggested that in this case the zero mode of $\varphi$
is a target time $t$. The requirement of renormalizability of the
world sheet $\sigma$ model implies that for the density matrix $\rho_{M}$
of a string state propagating in a background $\left\{ g_{i}\right\} $
\begin{equation}
\frac{d}{dt}\rho_{M}(g^{I},\, p_{I},t)=0\label{density}\end{equation}
 where $p_{i}$ is the conjugate momentum to $g^{i}$ within the framework
of a dynamical system with hamiltonian $H$ and action the Zamolodchikov
c-function \cite{Mavromatos:2009pp}, i.e. \begin{equation}
c[g]=\int dt(p_{I}\dot{g}^{I}-H).\label{dynamical}\end{equation}
 From (\textbackslash{}ref\{density\}) we deduce that \begin{equation}
\frac{\partial\rho}{\partial t}+\dot{g}^{I}\frac{\partial\rho}{\partial g^{I}}+\dot{p}_{I}\frac{\partial\rho}{\partial p^{I}}=0.\label{density2}\end{equation}
 The piece $\dot{p}_{I}\frac{\partial\rho}{\partial p^{I}}$ in (\ref{density2})
can be written as $G_{IJ}\beta^{J}\frac{\partial\rho}{\partial p^{I}}$;
using the canonical relationship of $g^{I}$ and $p^{I}$ this can
be recast as $-iG_{IJ}\beta^{J}[\rho,g^{J}]$. As discussed in \cite{Mavromatos:2009pp},
such a term leads to a non-unitary evolution of $\rho$. This analysis
cannot be regarded as being conclusive concerning the issue of non-unitarity,
even within the framework of non-critical string theory, since there
are issues relating to the time like signature of the Liouville field
and the identification of the local renormalization group scale. However
we will asssume that these caveats are not sufficient to invalidate
the main consequence and we shall explore the implications for $CPT\left(\equiv\Theta\right)$
symmetry where where the operators $C$, $P$, and $T$ denote charge
conjugation, parity and time reversal respectively.

\subsection{CPT }

Currently the successful theories are based on local Lorentz invariant
lagrangians, and it has been shown given the spin-statistics connection
that $CPT$ is a symmetry. The symmetry implies that the solution
set of a theory is invariant under reversal of parity, time and interchange
of particle and antiparticle. Consequently any violations of the consequences
of $CPT$ symmetry \cite{streater} would entail physics beyond the
standard model of particle physics which is based on lagrangians.
Typically the consequences of $CPT$ that are considered are those
of equal masses and lifetimes for particles and antiparticles. Recently
it was noted that when the $CPT$ operator is not well defined there
are implications for the symmetry structure of the initial entangled
state of two neutral mesons in meson factories such as DA$\Phi$NE,
the Frascati $\phi$ factory. Indeed, if $CPT$ can be defined as
a quantum mechanical operator, then the decay of a (generic) meson
with quantum numbers $J^{PC}=1^{--}$ \cite{Lipkin}, leads to a pair
state of neutral mesons $\left\vert i\right\rangle $ having the form
of the entangled state

\begin{equation}
\left\vert i\right\rangle =\frac{1}{\sqrt{2}}\left(\left\vert \overline{M_{0}}\left(\overrightarrow{k}\right)\right\rangle \left\vert M_{0}\left(-\overrightarrow{k}\right)\right\rangle -\left\vert M_{0}\left(\overrightarrow{k}\right)\right\rangle \left\vert \overline{M_{0}}\left(-\overrightarrow{k}\right)\right\rangle \right).\label{CPTV}\end{equation}
 This state has the Bose symmetry associated with particle-antiparticle
indistinguishability $C\mathcal{P}=+$, where $C$ is the charge conjugation
and $\mathcal{P}$ is the permutation operation. If, however, $CPT$
is not a good symmetry (i.e. ill-defined due to space-time foam),
then $M_{0}$ and $\overline{M_{0}}$ may not be identified and the
requirement of $C\mathcal{P}$ is relaxed\textasciitilde{}\cite{bernabeu}.
Consequently, in a perturbative framework, the state of the meson
pair can be parametrised to have the following form:\[
\left\vert i\right\rangle =\frac{1}{\sqrt{2}}\left(\left\vert \overline{M_{0}}\left(\overrightarrow{k}\right)\right\rangle \left\vert M_{0}\left(-\overrightarrow{k}\right)\right\rangle -\left\vert M_{0}\left(\overrightarrow{k}\right)\right\rangle \left\vert \overline{M_{0}}\left(-\overrightarrow{k}\right)\right\rangle \right)+\frac{\omega}{\sqrt{2}}\left|\Delta\left(\vec{k}\right)\right\rangle \]
 where \[
\left|\Delta\left(\vec{k}\right)\right\rangle \equiv\left|\overline{M}\left(\vec{k}\right)\right\rangle \left|M_{0}\left(-\vec{k}\right)\right\rangle +\left|M_{0}\left(\vec{k}\right)\right\rangle \left|\overline{M}\left(-\vec{k}\right)\right\rangle \]
 and $\omega=\left\vert \omega\right\vert e^{i\Omega}$ is a complex
$CPT$ violating (CPTV) parameter. For definiteness in what follows
we shall term this quantum-gravity effect in the initial state \cite{bernabeu}.

Decoherence or non-unitary evolution leads to a new method of breaking
$\Theta$. Unitary time evolution is an implicit assumption in the
proof of the $CPT$ theorem; Wald argued that non-unitary evolution
would allow $\Theta$ not to be a symmetry.Within the context of scattering
theory let $\mathcal{H}_{in}$ denote the Hilbert space of in states,
and $\bar{\mathcal{H}}_{in}$ the dual space. We can define the analogous
entities for the out states. Let $S$ be a mapping from the set of
in-states $\mathcal{G}_{in}$ to the set of out-states $\mathcal{G}_{out}$.
In our framework $\mathcal{G}_{in}$ is isomorphic to $\mathcal{H}\bigotimes\bar{\mathcal{H}}_{in}$
since the states are represented as density matrices $\rho_{B}^{A}$
where $A$ is a vector index associated with $\mathcal{H}_{in}$ and
$B$ is a vector index associated with $\bar{\mathcal{H}}_{in}$.
The indexed form of $S$ is $S_{b}^{a}{}_{C}^{D}$ where the lower
case indices refer to $\mathcal{H}_{out}$ and $\bar{\mathcal{H}}_{out}$.
If probability is conserved \begin{equation}
tr\left(S\rho\right)=\left(\rho\right)\label{prob}\end{equation}
 which in index notation can be written as $S_{a}^{a}{}_{C}^{D}=\delta_{C}^{D}$
where we have adopted the summation convention for repeated indices.
Consider now operators $\Theta_{in}$ and $\Theta_{out}$ which implement
the $CPT$ transformation on $\mathcal{G}_{in}$ and $\mathcal{G}_{out}$
respectively i.e.

\begin{eqnarray}
\Theta_{in} & \colon & \mathcal{G}_{in}\rightarrow\mathcal{G}_{out}\\
\Theta_{out} & \colon & \mathcal{G}_{out}\rightarrow\mathcal{G}_{in}.\end{eqnarray}
 In particular under a $CPT$ transformation let $\rho_{in}\;\epsilon\;\mathcal{G}_{in}$
be mapped into $\rho_{out}'$ and $\rho_{out}\;\epsilon\;\mathcal{G}_{out}$
be mapped into $\rho_{out}'$. If the theory (including quantum gravity)
is assumed to be invariant under $CPT$ then

\begin{eqnarray}
\rho_{out} & = & S\rho_{in},\label{CPT1}\\
\rho_{out}' & = & S\rho_{in}'\label{CPT2}\\
\Theta_{in}\rho_{in} & = & \rho_{out}',\label{scat1}\\
\Theta_{out}\rho_{out} & = & \rho_{in}',\label{scat2}\end{eqnarray}
 and \begin{equation}
\Theta_{in}\Theta_{out}=I,\;\Theta_{out}\Theta_{in}=I.\label{compos}\end{equation}

Since \begin{equation}
\Theta_{out}=\Theta_{in}^{-1}\end{equation}
 it is convenient to drop the suffix $in$ in this last relation.
Hence from (\ref{CPT1}), (\ref{CPT2}),(\ref{scat1})and (\ref{scat2})we
can deduce that \begin{eqnarray}
\Theta\rho_{in} & = & \rho_{out}',\label{scat3}\\
 & = & S\rho_{in}'\\
 & = & S\Theta^{-1}\rho_{out}\\
 & = & S\Theta^{-1}S\rho_{in},.\label{inv1}\end{eqnarray}
 From (\ref{scat3}) and (\ref{inv1}) we can deduce the important
result that $S$ has an inverse given by $\Theta^{-1}S\Theta^{-1}$.
Hence if $\Theta$exists then time reversed evolution is permitted.
Consequently for non-unitary evolution $\Theta$ cannot be defined.

Another possible contribution of breakdown of $CPT$ invariance is
towards the generation of cosmological charge asymmetry (baryogenesis).
This would provide an additional mechanism to that proposed by Sakhorov
based on: non-conservation of baryon number; deviation from thermal
equilibrium; and $C$ and $CP$ violation. This has, in the past,
been addressed within thermal equilibrium by using a mass difference
between particle and anti-particle, a possible consequence of $CPT$
violation. In our case of D-foam the decoherence effects on particle
and anti-particle may be different, a fact we exploit.

\section{Effective model of D-foam}

For an observer on the brane world the crossing D-particles will appear
as twinkling space-time defects, i.e. microscopic space-time fluctuations.
This will give the four-dimensional brane world a {}``D-foamy''
structure. In phase space, for a D3-brane world, the function $u_{i}$,
involving a momentum transfer, $\Delta k_{i}$, can be modelled by
a local operator using the following parametrization~\cite{Dparticle}:
\begin{equation}
u_{i}=g_{s}\frac{\Delta k_{i}}{2M_{s}}=r_{i}k_{i}\,~,\,{\rm no~sum}\, i=1,2,3~,\label{defu2}\end{equation}
 where the (dimensionful) variables $r_{i},i=1,2,3$, appearing above,
are related to the fraction of momentum that is transferred at a collision
with a D-particle in each spatial direction $i$ .The target space-time
metric state, which is close to being flat, can be represented schematically
as a density matrix

\begin{equation}
\rho_{\mathrm{grav}}=\int d\,^{4}\tilde{r}\,\, f\left(\tilde{r}_{\mu}\right)\left|g\left(\tilde{r}_{\mu}\right)\right\rangle \left\langle g\left(\tilde{r}_{\mu}\,\right)\right|.\,\label{gravdensity}\end{equation}
 The parameters $\tilde{r}_{\mu}\,\left(\mu=0,\ldots,4\right)$ are
stochastic with a gaussian distribution $\, f\left(\widetilde{r}_{\mu}\,\right)$
characterised by the averages\%

\begin{equation}
\left\langle \tilde{r}_{\mu}\right\rangle =0,\;\left\langle \tilde{r}_{\mu}\tilde{r}_{\nu}\right\rangle =\Delta_{\mu}\delta_{\mu\nu}\,.\label{stoch}\end{equation}
 The fluctuations experienced by the two entangled neutral mesons
will be assumed to be independent and $\Delta_{\mu}\sim O\left(\frac{E^{2}}{M_{P}^{2}}\right)$i.e.
very small. As matter moves through the space-time foam, assuming
ergodicity, the effect of time averaging is assumed to be equivalent
to an ensemble average. As far as our present discussion is concerned
we will consider a semi-classical picture for the metric and so $\left|g\left(\tilde{r}_{\mu}\right)\right\rangle $
in (\ref{gravdensity}) will be a coherent state. In order to address
two {}``flavours'' the fluctuations of each component of the metric
tensor $g^{\alpha\beta}$ will not be simply given by the simple recoil
distortion (\ref{recoil}), but instead will be taken to have a $2\times2$
({}``flavour'') structure:

\begin{eqnarray}
g_{00} & = & -\mathsf{1},\nonumber \\
g_{01} & = & g_{10}=\widetilde{r}_{0}\mathsf{\sigma_{0}}+\tilde{r}_{1}\sigma_{1}+\widetilde{r}_{2}\sigma_{2}\label{matansatz}\\
g_{jj} & = & 1\end{eqnarray}
 where $\sigma_{0}=1_{2},($the identity matrix). The modelling of
the metric can be made more elaborate, but the salient feature is
the structure of $g^{01}$( where for simplicity we have taken the
momentum to be in the 1-direction). For the case of the omega effect
this restriction is acceptable since the K-mesons are produced collinearly
in a $\phi$-meson factory (in the centre of mass frame).A more general
ansatz than (\ref{matansatz}) (which will reduce to this when momenta
are collinear) has the form \[
g_{0j}=r_{\mu}\sigma_{\mu}\hat{k_{j}}\]
 where $\hat{k_{j}}$ is the momentum operator of a particle moving
in this background; this is the Finsler nature of the metric.We use
the term flavour in a general sense. For K mesons flavour would mean
$K_{L}$or $K_{S}$. An elementary D-particle will typically affect
neutral particles because it does not carry charge and a charge string
attched to it would not have anywhere for the flux to flow. Elementary
D-particles appear in some string theories such as $\mathrm{IIa}$
but in other more phenomenologically relevant string theories they
do not. In such cases, effective D-particle behaviour can occur from
D3 branes wrapped around 3-cycles. Consequently such D-particles could
interact with charged strings and hence contribute to baryogenesis.
However the treatment of D-particles, arising from compactifications,
is technically more involved and the orders of magnitude will depend
on the details of the compactification process. We nonetheless will
point out how the D-foam formally can give rise to particle-antiparticle
asymmetry.

For the neutral Kaon system, the case of interest, $K_{0}-\overline{K}_{0}$,
is produced by a $\phi$-meson at rest,i.e. $K_{0}-\overline{K}_{0}$
in their C.M. frame. The CP eigenstates (on choosing a suitable phase
convention for the states $\left|K_{0}\right\rangle $ and $\left|\overline{K_{0}}\right\rangle $
) are, in standard notation, $\left|K_{\pm}\right\rangle $ with

\[
\left|K_{\pm}\right\rangle =\frac{1}{\sqrt{2}}\left(\left|K_{0}\right\rangle \pm\left|\overline{K_{0}}\right\rangle \right).\]
 The mass eigensates $\left|K_{S}\right\rangle $ and $\left|K_{L}\right\rangle $
are written in terms of $\left|K_{\pm}\right\rangle $ as\[
\left|K_{L}\right\rangle =\frac{1}{\sqrt{1+\left|\varepsilon_{2}\right|^{2}}}\left[\left|K_{-}\right\rangle \,+\varepsilon_{2}\left|K_{+}\right\rangle \right]\]
 and \[
\left|K_{S}\right\rangle =\frac{1}{\sqrt{1+\left|\varepsilon_{1}\right|^{2}}}\left[\left|K_{+}\right\rangle \,+\varepsilon_{1}\left|K_{+}\right\rangle \right].\]
 In terms of the mass eigenstates \[
\left|i\right\rangle =\mathcal{C}\left\{ \begin{array}{c}
\left(\left|K_{L}\left(\overrightarrow{k}\right)\right\rangle \left|K_{S}\left(-\overrightarrow{k}\right)\right\rangle -\left|K_{S}\left(\overrightarrow{k}\right)\right\rangle \left|K_{L}\left(-\overrightarrow{k}\right)\right\rangle \right)+\\
\omega\left(\left|K_{S}\left(\overrightarrow{k}\right)\right\rangle \left|K_{S}\left(-\overrightarrow{k}\right)\right\rangle -\left|K_{L}\left(\overrightarrow{k}\right)\right\rangle \left|K_{L}\left(-\overrightarrow{k}\right)\right\rangle \right)\end{array}\right\} \]
 where $\mathcal{C=}\frac{\sqrt{\left(1+\left|\varepsilon_{1}\right|^{2}\right)\left(1+\left|\varepsilon_{2}\right|^{2}\right)}}{\sqrt{2}\left(1-\varepsilon_{1}\varepsilon_{2}\right)}$
\cite{bernabeu}. In the notation of two level systems (on suppressing
the $\overrightarrow{k}$ label) we write

\begin{eqnarray}
\left|K_{L}\right\rangle  & = & \left|\uparrow\right\rangle \\
\left|K_{S}\right\rangle  & = & \left|\downarrow\right\rangle .\end{eqnarray}
 These will be our {}``flavours'' and represent the two physical
eigenstates, with masses $m_{1}\equiv m_{L}$, $m_{2}\equiv m_{S}$,
with

\begin{equation}
\Delta m=m_{L}-m_{S}\sim3.48\times10^{-15}~\mathrm{GeV}~.\label{deltam}\end{equation}

The Klein-Gordon equation for a spinless neutral meson field $\Phi=\left(\begin{array}{c}
\phi_{1}\\
\phi_{2}\end{array}\right)$ with mass matrix $m=\frac{1}{2}\left(m_{1}+m_{2}\right)\mathsf{1}+$
$\frac{1}{2}\left(m_{1}-m_{2}\right)\sigma_{3}$ in a gravitational
background is

\begin{equation}
(g^{\alpha\beta}D_{\alpha}D_{\beta}-m^{2})\Phi=0\label{KleinGordon}\end{equation}
 where $D_{\alpha}$ is the covariant derivative. Since the Christoffel
symbols vanish for $a_{i}$ independent of space time the $D_{\alpha}$
coincide with $\partial_{\alpha}$. Hence \begin{equation}
\left(g^{00}\partial_{0}^{2}+2g^{01}\partial_{0}\partial_{1}+g^{11}\partial_{1}^{2}\right)\Phi-m^{2}\Phi=0.\label{KG2}\end{equation}
 It is useful at this stage to rewrite the state $\left|i\right\rangle $
in terms of the mass eigenstates.

The unnormalised state $\left|i\right\rangle $ will then be an example
of an initial state

\begin{equation}
\left|\psi\right\rangle =\left|k,\uparrow\right\rangle ^{\left(1\right)}\left|-k,\downarrow\right\rangle ^{\left(2\right)}-\left|k,\downarrow\right\rangle ^{\left(1\right)}\left|-k,\uparrow\right\rangle ^{\left(2\right)}+|\Delta\rangle\end{equation}
 with

\[
|\Delta\rangle=\xi\left|k,\uparrow\right\rangle ^{\left(1\right)}\left|-k,\uparrow\right\rangle ^{\left(2\right)}+\xi^{\prime}\left|k,\downarrow\right\rangle ^{\left(1\right)}\left|-k,\downarrow\right\rangle ^{\left(2\right)}\]
 where $\left|K_{L}\left(\overrightarrow{k}\right)\right\rangle =\left|k,\uparrow\right\rangle $
and we have taken $\overrightarrow{k}$ to have only a non-zero component
$k$ in the $x$-direction; superscripts label the two separated detectors
of the collinear meson pair, $\xi$ and $\xi^{\prime}$ are complex
constants and we have left the state $\left|\psi\right\rangle $ unnormalised.
The evolution of this state is governed by a hamiltonian $\widehat{H}$

\begin{equation}
\widehat{H}=g^{01}\left(g^{00}\right)^{-1}\widehat{k}-\left(g^{00}\right)^{-1}\sqrt{\left(g^{01}\right)^{2}{k}^{2}-g^{00}\left(g^{11}k^{2}+m^{2}\right)}\label{GenKG}\end{equation}
 which is the natural generalisation of the standard Klein-Gordon
hamiltonian in a one particle situation. Moreover $\widehat{k}\left|\pm k,\uparrow\right\rangle =\pm k\left|k,\uparrow\right\rangle $
together with the corresponding relation for $\downarrow$. We next
note that the Hamiltonian interaction terms

\begin{equation}
\widehat{H_{I}}=-\left(r_{1}\sigma_{1}+r_{2}\sigma_{2}\right)\widehat{k}\label{inter}\end{equation}
 are the leading order contribution in the small parameters $r_{\mu}$
in the Hamiltonian $H$ (\ref{GenKG}), since the corresponding variances
$\sqrt{\Delta_{\mu}}$ are small. The term (\ref{inter}), has been
used in \cite{bernabeu} as a perturbation in the framework of non-degenerate
perturbation theory, in order to derive the {}``gravitationally-dressed''
initial entangled meson states, immediately after the $\phi$ decay.
The result is:

\begin{equation}
\left\vert k,\uparrow\right\rangle _{QG}^{\left(1\right)}\left\vert -k,\downarrow\right\rangle _{QG}^{\left(2\right)}-\left\vert k,\downarrow\right\rangle _{QG}^{\left(1\right)}\left\vert -k,\uparrow\right\rangle _{QG}^{\left(2\right)}=|\Sigma\rangle+|\widetilde{\Delta}\rangle\end{equation}
 where

\[
|\Sigma\rangle=\left\vert k,\uparrow\right\rangle ^{\left(1\right)}\left\vert -k,\downarrow\right\rangle ^{\left(2\right)}-\left\vert k,\downarrow\right\rangle ^{\left(1\right)}\left\vert -k,\uparrow\right\rangle ^{\left(2\right)},\]

\[
|\widetilde{\Delta}\rangle=\begin{array}{c}
\left\vert k,\downarrow\right\rangle ^{\left(1\right)}\left\vert -k,\downarrow\right\rangle ^{\left(2\right)}\left(\beta^{\left(1\right)}-\beta^{\left(2\right)}\right)+\left\vert k,\uparrow\right\rangle ^{\left(1\right)}\left\vert -k,\uparrow\right\rangle ^{\left(2\right)}\left(\alpha^{\left(2\right)}-\alpha^{\left(1\right)}\right)\\
+\beta^{\left(1\right)}\alpha^{\left(2\right)}\left\vert k,\downarrow\right\rangle ^{\left(1\right)}\left\vert -k,\uparrow\right\rangle ^{\left(2\right)}-\alpha^{\left(1\right)}\beta^{\left(2\right)}\left\vert k,\uparrow\right\rangle ^{\left(1\right)}\left\vert -k,\downarrow\right\rangle ^{\left(2\right)}\end{array}\]
 and

\begin{equation}
\alpha^{\left(i\right)}=\frac{^{\left(i\right)}\left\langle \uparrow,k^{\left(i\right)}\right\vert \widehat{H_{I}}\left\vert k^{\left(i\right)},\downarrow\right\rangle ^{\left(i\right)}}{E_{2}-E_{1}}~,\quad\beta^{\left(i\right)}=\frac{^{\left(i\right)}\left\langle \downarrow,k^{\left(i\right)}\right\vert \widehat{H_{I}}\left\vert k^{\left(i\right)},\uparrow\right\rangle ^{\left(i\right)}}{E_{1}-E_{2}}~,~\quad i=1,2\label{qgpert2}\end{equation}
 where the index $(i)$ runs over meson species ({}``flavours'')
($1\rightarrow K_{L},~2\rightarrow K_{S}$). The reader should notice
that the terms proportional to $\left(\alpha^{\left(2\right)}-\alpha^{\left(1\right)}\right)$
and $\left(\beta^{\left(1\right)}-\beta^{\left(2\right)}\right)$
in (\textbackslash{}ref\{entangl\}) generate $\omega$-like effects.
We concentrate here for brevity and concreteness in the strangeness
conserving case of the $\omega$-effect in the initial decay of the
$\phi$ meson \cite{bernabeu}, which corresponds to $r_{i}\propto\delta_{i2}$.
We should mention, however, that in general quantum gravity does not
have to conserve this quantum number, and in fact strangeness-violating
$\omega$-like terms are generated in this problem through time evolution
\cite{bernabeu}.

We next remark that, on averaging the density matrix over the random
variables $r_{i}$, which are treated as independent variables between
the two meson particles of the initial state , we observe that only
terms of order $|\omega|^{2}$ will survive, with the order of $|\omega|^{2}$
being

\begin{eqnarray}
|\omega|^{2} & = & \widetilde{\Delta}_{(1),(2)}\left(\mathcal{O}\left(\frac{1}{(E_{1}-E_{2})}(\langle\downarrow,k|H_{I}|k,\uparrow\rangle)^{2}\right)\right),\nonumber \\
 & = & \widetilde{\Delta}_{(1),(2)}\left(\mathcal{O}\left(\frac{\Delta_{2}k^{2}}{(E_{1}-E_{2})^{2}}\right)\right)\sim\widetilde{\Delta}_{(1),(2)}\left(\frac{\Delta_{2}k^{2}}{(m_{1}-m_{2})^{2}}\right)\label{omegaorder}\end{eqnarray}
 for the physically interesting case of non-relativistic Kaons in
$\phi$ factories, in which the momenta are of order of the rest energies.
The notation $\widetilde{\Delta}_{(1),(2)}\left(\dots\right)$ above
indicates that one considers the differences of the variances $\Delta_{2}$
between the two mesons $1$ - $2$, in that order.

The variances in our model of D-foam, which are due to quantum fluctuations
of the recoil velocity variables about the zero average (dictated
by the imposed requirement on Lorentz invariance of the string vacuum)
lead for the square of the amplitude of the (complex) $\omega$-parameter:

\begin{equation}
|\omega|^{2}\sim g_{0}^{2}\frac{\left(m_{1}^{2}-m_{2}^{2}\right)}{M_{s}^{2}}\frac{k^{2}}{(m_{1}-m_{2})^{2}}=\frac{m_{1}+m_{2}}{m_{1}-m_{2}}~\frac{k^{2}}{(M_{s}^{2}/g_{0}^{2})},\label{final}\end{equation}
 where $M_{P}\equiv M_{s}/g_{0}$ represents the (average) quantum
gravity scale, which may be taken to be the four-dimensional Planck
scale. In general, $M_{s}/g_{0}$ is the (average) D-particle mass,
as already mentioned. In the modern version of string theory, $M_{s}$
is arbitrary and can be as low as a few TeV, but in order to have
phenomenologically correct string models with large extra dimensions
one also has to have in such cases very weak string couplings $g_{0}$,
such that even in such cases of low $M_{s}$, the D-particle mass
$M_{s}/g_{0}$ is always close to the Planck scale $10^{19}$ GeV.
But of course one has to keep an open mind about ways out of this
pattern, especially in view of the string landscape.

The result (\ref{final}), implies, for neutral Kaons in a $\phi$
factory, for which (\textbackslash{}ref\{deltam\}) is valid, a value
of: $|\omega|\sim10^{-11}$, which in the sensitive $\eta^{+-}$ bi-pion
decay channel, is enhanced by three orders of magnitude, as a result
of the fact that the $|\omega|$ effect always appears in the corresponding
observables \cite{bernabeu} in the form $|\omega|/|\eta^{+-}|$,
and the CP-violating parameter $|\eta^{+-}|\sim10^{-3}$. Unfortunately,
this value is still some two orders of magnitude away from current
bounds of the $\omega$-effect at, or the projected sensitivity of
upgrades of, the DA$\Phi$NE detector.

A similar calculation can be done for the particle-antiparticle asymmetry
except that we will use the more general metric resulting in \begin{equation}
H=E_{r}\left(\overrightarrow{k}\right)1_{2}+r_{\mu}\sigma_{\mu}\left(\overrightarrow{k}\right)^{2}\label{asymmetry}\end{equation}
where \[
E_{r}\left(\overrightarrow{k}\right)=\sqrt{m^{2}+\sum_{j}k_{j}^{2}+\left(r_{\mu}\right)^{2}\left(\sum_{j}k_{j}^{2}\right)^{2}-\sum r_{j}^{2}k_{j}^{4}}\]
 This leads to gravitational dressing of the particle and antiparticle
states so that the masses get shifted so as to induce a mass difference;
in thermodyanmic equilibrium the canonical number distribution ( Bose-Einstein
or Fermi-Direc) would give a difference between the particle and anti-particle
number. Explcitly, if the eigenvalues of $H$ are denoted by $a_{1}\left(\overrightarrow{k},r_{\mu}\right)$
and $a_{2}\left(\overrightarrow{k},r_{\mu}\right)$ then to lowest
order we can show that $a_{2}\left(\overrightarrow{k},r_{\mu}\right)-a_{1}\left(\overrightarrow{k},r_{\mu}\right)=2$$\left|\overrightarrow{k}\right|\left|\overrightarrow{r}\right|$
and clearly there will then be a difference between the grand canonical
number distribution functions \[
n_{a_{j}}\left(\overrightarrow{k},r_{\mu},\xi\right)=\frac{1}{\exp\left(\beta\left[a_{j}\left(\overrightarrow{k},r_{\mu}\right)-\mu\right]\right)+\xi},\:\; j=1,2\]
 where $\xi=\pm1$ for different $j$.

\section*{Conclusions}

In our D-foam model, string matter on a brane world interacts with
D-particles in the bulk. Recoil of the heavy D particles owing to
interactions with the stringy matter produces a gravitational distortion
which has a back-reaction on matter. There is information loss at
the recoil which leads to non-unitary evolution for the matter. From
the general considerations of Wald, CPT symmetry may then be violated.
This new mechanism of violation is explored for the omega effect as
well as matter-anti-matter asymmetry. For the omega effect there is
a chance that the upgrade to the the DA$\Phi NE$ detector will allow
a stringent test of the D-foam prediction. Particle-antiparticle asymmetry
has more theoretical uncertainities in D-foam predictions and further
investigations are needed using D-brane instantons for example.

\section*{Acknowledgements}

My work is partially supported by the European Union through the Marie
Curie Research and Training Network \emph{UniverseNet} (MRTN-2006-035863).
The work presented in this paper has arisen due to discussion as well
as collaboration with N.E. Mavromatos.

\section*{Appendix A: Calculation of back reaction in D-particle foam}

\subsection*{D-particle foam contributions to master equation for Liouville-decoherence}

The material in this Appendix is a review based on \cite{kmw}, where
we refer the reader for further details. Let us consider a $D$-particle,
located at $y^{i}(t=0)\equiv y_{i}$ of the spatial coordinates of
a $(d+1)$-dimensional space time (which could be a D3-brane world),
which at a time $t=0$ experiences an impulse, as a result of scattering
with a matter string state . In a $\sigma$-model framework, the trajectory
of the $D$-particle $y^{i}(t)$, $i=1,2,\dots d,$ a spatial index,
is described by inserting the following vertex operator in the $\sigma$-model
of a free string: \begin{equation}
V=\int_{\partial\Sigma}g_{ij}y^{j}(t)\partial_{n}X^{i}\label{path1}\end{equation}
 where $g_{ij}$ denotes the spatial components of the metric, $\partial\Sigma$
denotes the world-sheet boundary, $\partial_{n}$ is a normal world-sheet
derivative, $X^{i}$ are $\sigma$-model fields obeying Dirichlet
boundary conditions on the world sheet, and $t$ is a $\sigma$-model
field obeying Neumann boundary conditions on the world sheet, whose
zero mode is the target time. The space-time prior to Liouville dressing
is assumed Euclidean for formal reasons (convergence of the corresponding
$\sigma$-model path integral). We note, however, that the final Liouville-dressed
target space-time acquires Minkowski signature as a result of the
time-like signature of the Liouville mode~\cite{gravanis}.

In the non-relativistic approximation, appropriate for a heavy D-particle
defect of mass $M_{s}/g_{s}$, with $M_{s}$ the string scale, and
$g_{s}$ the string coupling, assumed weak ($g_{s}\ll1$), the path
$y^{i}(t)$ corresponding to the impulse is given by: \begin{eqnarray}
y_{i}(t) & = & \left(\varepsilon y_{i}+u_{i}t\right)\Theta_{\varepsilon}(t)\label{path}\\
u_{i} & = & \left(k_{1}-k_{2}\right)_{i}~,\end{eqnarray}
 with $k_{1}\left(k_{2}\right)$ the momentum of the propagating string
state before (after) the recoil $y_{i}$ are the spatial collective
coordinates of the D particle, and the regularized Heaviside functional
operator $\Theta_{\varepsilon}(t)$ is given by (\ref{heaviside})
in the text~\cite{kmw}: \begin{eqnarray}
\Theta_{\varepsilon}\left(t\right) & = & \frac{1}{2\pi i}\int_{-\infty}^{\infty}\frac{dq}{q-i\varepsilon}e^{iqt},\label{heaviside2}\end{eqnarray}
 Eq. (\ref{path}) contains actually a \emph{pair} of deformations
corresponding to the $\sigma$-model couplings $y_{i}$ and $u_{i}$.
These deformations are relevant in a world-sheet renormalization-group
sense, having anomalous scaling dimension $-\frac{\varepsilon^{2}}{2}$,
i.e. to leading order in a coupling constant expansion their renormalization-group
$\beta$-functions read: \begin{equation}
\beta^{y^{i}}=-\frac{\varepsilon^{2}}{2}y^{i}~,\qquad\beta^{u^{i}}=-\frac{\varepsilon^{2}}{2}u^{i}~.\label{betafnct}\end{equation}
 The deformations form a logarithmic conformal algebra (superconformal
algebra in the case of superstrings) which \emph{closes} if and only
if one identifies~\cite{kmw} the regulating parameter $\varepsilon^{-2}$
with the world-sheet renormalization-group scale $\mathrm{ln}|L/a|^{2}$
($L(a)$ is the Infrared (Ultraviolet) world-sheet scale): \begin{equation}
\varepsilon^{-2}=\eta\mathrm{ln}|L/a|^{2}\label{elog}\end{equation}
 where $\eta$ denotes the signature of time $t$ of the target-space
manifold of the $\sigma$-model (prior to Liouville dressing). For
Euclidean manifolds, assumed here for path-integral convergence, $\eta=+1$.

Upon the identification (\ref{elog}) the re-scaled couplings $\overline{y}_{i}\equiv\frac{y_{i}}{\varepsilon}$
and $\overline{u}_{i}\equiv\frac{u_{i}}{\varepsilon}$ are \emph{marginal},
that is independent of the scale $\varepsilon$. It is these marginal
couplings that are connected to target-space quantities of physical
significance, such as the space-time back reaction of recoil. In the
limit $\varepsilon\to0$ ( the long-time limit) the dominant contributions
come from the $u_{i}$ recoil deformation in (\ref{path}).


\begin{thebibliography}{17}
\bibitem{finsler} see, for instance: D. Bao, S.~S.~Chern and Z.~Shen,
\emph{An introduction to Finsler Geometry} (Springer-Verlag (NY, 2000)); For a recent review, with applications to cosmology, see S. Vacaru, {}``Principles of Einstein-Finsler Gravity and Cosmology,''
arXiv: 1012. 4148 [physics.gen-ph].


\bibitem{polch2}J. Polchinski,{}``Dirichlet branes and Ramond-Ramond
charges'', Phys. Rev. Lett. 75 4724 (1995).

\bibitem{emw} J. R. Ellis, N. E. Mavromatos and M.Westmuckett, {}``A
supersymmetric D-brane model of space-time foam,'' Phys. Rev. D 70,
044036 (2004) {[}arXiv:gr-qc/0405066{]}; {}``Potentials between D-branes
in a supersymmetric model of space-time foam'', Phys.Rev. D 71, 106006
(2005) {[}arXiv:gr-qc/0501060{]}.

\bibitem{kmw} I. I. Kogan, N. E. Mavromatos and J. F. Wheater, {}``D-brane
recoil and logarithmic operators,'' Physics Letters B 387, 483 (1996)
{[}arXiv:hep-th/9606102{]}; For a review focusing on D-brane recoil,
including supermembranes, see: N. E. Mavromatos, {}``Logarithmic
conformal field theories and strings in changing backgrounds,'' arXiv:hep-th/0407026,
in Shifman, M. (ed.) et al.: From fields to strings, I. Kogan memorial
Volume 2\}, 1257-1364. (World Sci. 2005), and references therein.

\bibitem{Anchordoqui:2006xv} L. A. Anchordoqui, {}``Spacetime foam
at a TeV,'' J. Phys.Conf. Ser. 60 (2007) 191 {[}arXiv:hep-ph/0610025{]}.

\bibitem{ArkaniHamed:1998rs} N. Arkani-Hamed, S. Dimopoulos and G.
R. Dvali, {}``The hierarchy problem and new dimensions at a millimeter,''
Phys. Lett. B 429, 263 (1998) {[}arXiv:hep-ph/9803315{]}.

\bibitem{Dfoam} J.~R.~Ellis, N.~E.~Mavromatos and D.~V.~Nanopoulos,
Gen.\ Rel.\ Grav.\ \textbf{32}, 127 (2000) {[}arXiv:gr-qc/9904068{]};
 Phys.\ Rev.\ D \textbf{61}, 027503 (2000) {[}arXiv:gr-qc/9906029{]};
 Phys.\ Rev.\ D \textbf{62}, 084019 (2000) {[}arXiv:gr-qc/0006004{]}.
 J.~R.~Ellis, K.~Farakos, N.~E.~Mavromatos, V.~A.~Mitsou and
D.~V.~Nanopoulos, 
 Astrophys.\ J.\ \textbf{535}, 139 (2000) {[}arXiv:astro-ph/9907340{]};
 J.~R.~Ellis, N.~E.~Mavromatos, D.~V.~Nanopoulos and G.~Volkov,
 Gen.\ Rel.\ Grav.\ \textbf{32}, 1777 (2000) {[}arXiv:gr-qc/9911055{]}.
J.~R.~Ellis, N.~E.~Mavromatos and M.~Westmuckett, 
Phys.\ Rev.\ D \textbf{70}, 044036 (2004) {[}arXiv:gr-qc/0405066{]}.


\bibitem{shore}G. M. Shore,{}``A local renormalization group equation,
diffeomorphisms and conformal invariance in sigma-models'', Nuclear
Physics B \textbf{286} 349 (1987)

\bibitem{zam}A. B. Zamolodchikov, {}``'Irreversibility' Of The Flux
Of The Renormalization Group In A 2-D Field Theory,'' JETP Lett.
43, 730 (1986) {[}Pisma Zh. Eksp. Teor. Fiz.43, 565 (1986){]}.

\bibitem{gutperle}M. Gutperle, M. Headrick, S. Minwalla and M. Schomerus,
{}``Space-time energy decreases under world-sheet renormalization
group flow'', Journal of High Energy Physics, 1 073 (2003)

\bibitem{EllisErice}J. Ellis, N. E. Mavromatos and D. V. Nanopoulos,
{}``A non-critical string approach to black holes, time and quantum
dynamics'',in {}``From supersymmetry to the origin of space-time''
Ed A. Zichichi, World Scientific (1995), arXiv:hep-th/9403133

\bibitem{Mavromatos:2009pp} N. E. Mavromatos, {}``Decoherence and
CPT Violation in a Stringy Model of Space-Time Foam,'' Found.Phys.
40, 917 (2010) {[}arXiv:0906.2712 {[}hep-th{]}{]}.

\bibitem{streater} R. F. Streater and A. S. Wightman, PCT, Spin and
Statisitics, and All That, Benjamin, New York, (1964); O. W. Greenberg,
{}``Why is CPT fundamental?,'' Found. Phys. 36, 1535 (2006) {[}arXiv:hep-ph/0309309{]}.

\bibitem{Lipkin}H. J. Lipkin, {}``CP violation and coherent decays
of kaon pairs,'' Phys. Rev. 176 1715 (1968)

\bibitem{bernabeu}J. Bernabeu, N. E. Mavromatos and S. Sarkar, {}``Decoherence
induced CPT violation and entangled neutral mesons,'' Physical\textbackslash{}
Review D 74, 045014 (2006) {[}arXiv:hep-th/0606137{]}.

\bibitem{Dparticle} N.E.\textasciitilde{}Mavromatos and Sarben Sarkar,
{}``Liouville decoherence in a model of flavour oscillations in the
presence of dark energy,'' Physical Review D 72, 065016 (2005) {[}arXiv:hep-th/0506242{]}.

\bibitem{gravanis} E. Gravanis and N. E. Mavromatos, {}``Vacuum
energy and cosmological supersymmetry breaking in brane worlds,''
Physics Letters B 547, 117 (2002) {[}arXiv:hep-th/0205298{]}.
\end{thebibliography}
\end{document}